\documentclass[twoside]{sfm}

\input{sf.def}

\begin{document}

\def\llm{{\sc LLmodels}}
\def\atl{{\sc ATLAS9}}
\def\aatl{{\sc ATLAS12}}
\def\starsp{{\sc STARSP}}
\def\aur{$\Theta$~Aur}
\def\logg{\log g}
\def\tauros{\tau_{\rm Ross}}
\def\kms{km\,s$^{-1}$}
\def\bz{$\langle B_{\rm z} \rangle$}
\def\degr{^\circ}
\def\aaps{A\&AS}
\def\aap{A\&A}
\def\apjs{ApJS}
\def\apj{ApJ}
\def\rmxaa{Rev. Mexicana Astron. Astrofis.}
\def\mnras{MNRAS}
\def\actaa{Acta Astron.}
\newcommand{\Tef}{T$_{\rm eff}$~}
\newcommand{\Vt}{$V_t$}
\newcommand{\CC}{$^{12}$C/$^{13}$C~}
\newcommand{\CDC}{$^{12}$C/$^{13}$C~}

\pagebreak

\thispagestyle{titlehead}

\setcounter{section}{0}
\setcounter{figure}{0}
\setcounter{table}{0}

\markboth{Butkovskaya V.}{Magnetic field of Vega}

\titl{Magnetic field of Vega}{Butkovskaya V.}
{Crimean Astrophysical Observatory of Taras Shevchenko National University of Kyiv, Nauchny, Ukraine, email: {\tt varya@crao.crimea.ua}}
 
 \abstre{Recent spectropolarimetric observations of Vega have revealed a weak magnetic field. I present here the results of spectropolarimetric study of Vega during 37 nights from 1997 to 2012. 
}

\baselineskip 12pt

\section{Introduction}

Vega ($\alpha$ Lyr, A0V) is one of the brightest and most familiar stars in the night sky, termed by astronomers as 'arguably the next most important star in the sky after the Sun' (Gulliver et al. \cite{gulliver94}). Vega demonstrates a low projected rotational velocity ($v$sin$i$ < 22 km/s), but it is a rapidly rotating star seen nearly pole-on. The rapid rotation is causing the equator to bulge outward because of centrifugal effects, and, as a result, there is a latitude variation of temperature that reaches a maximum at the poles. 

No consensus has yet been reached as to how fast Vega is rotating. Different authors propose the different rotation period: 0.525 d (Aufdenberg et al. \cite{aufdenberg06}, interferometry), 0.662 d (Hill et al. \cite{hill10}, high resolution spectral line profiles), 0.732 d (Petit et al. \cite{petit10}, magnetic maps), 0.733 d (Takeda et al. \cite{takeda08}, modelling of individual spectral lines and the spectral energy distribution).

Butkovskaya et al. \cite{butk11} using own spectropolarimetric data have confirmed the 21-year variability of Vega discovered by Vasil'ev et al.  \cite{vasil89} from spectrophotometry. They found that the equivalent widths of the four spectral lines Mg I 5167.321, Mg I 5172.684, Mg I 5183.604, and Fe II 5169.033 vary with the 21-year period.

Recent spectropolarimetric observations of Vega have revealed the presence of a weak magnetic field of 0.6 $\pm$ 0.2 G (Lignieres et al. \cite{lignieres09}, Petit et al. \cite{petit10}). Alina et al. \cite{alina11}  noted the stability of the weak magnetic field on timescale of 3 years. A potentially significant fraction of A stars might display a similar type of magnetism, but the geometry of these fields is poorly constrained. To answer this question it is crucial to accumulate more information about surface magnetic structures. Therefore, the long-term spectropolarimetric observations of Vega are needed. We present here the results of spectropolarimetric study of Vega during 37 nights in 1997-2012.

\section{Observations}\label{obs}

\begin{figure}[!t]
\begin{center}
\hbox{
 \includegraphics[width=10cm]{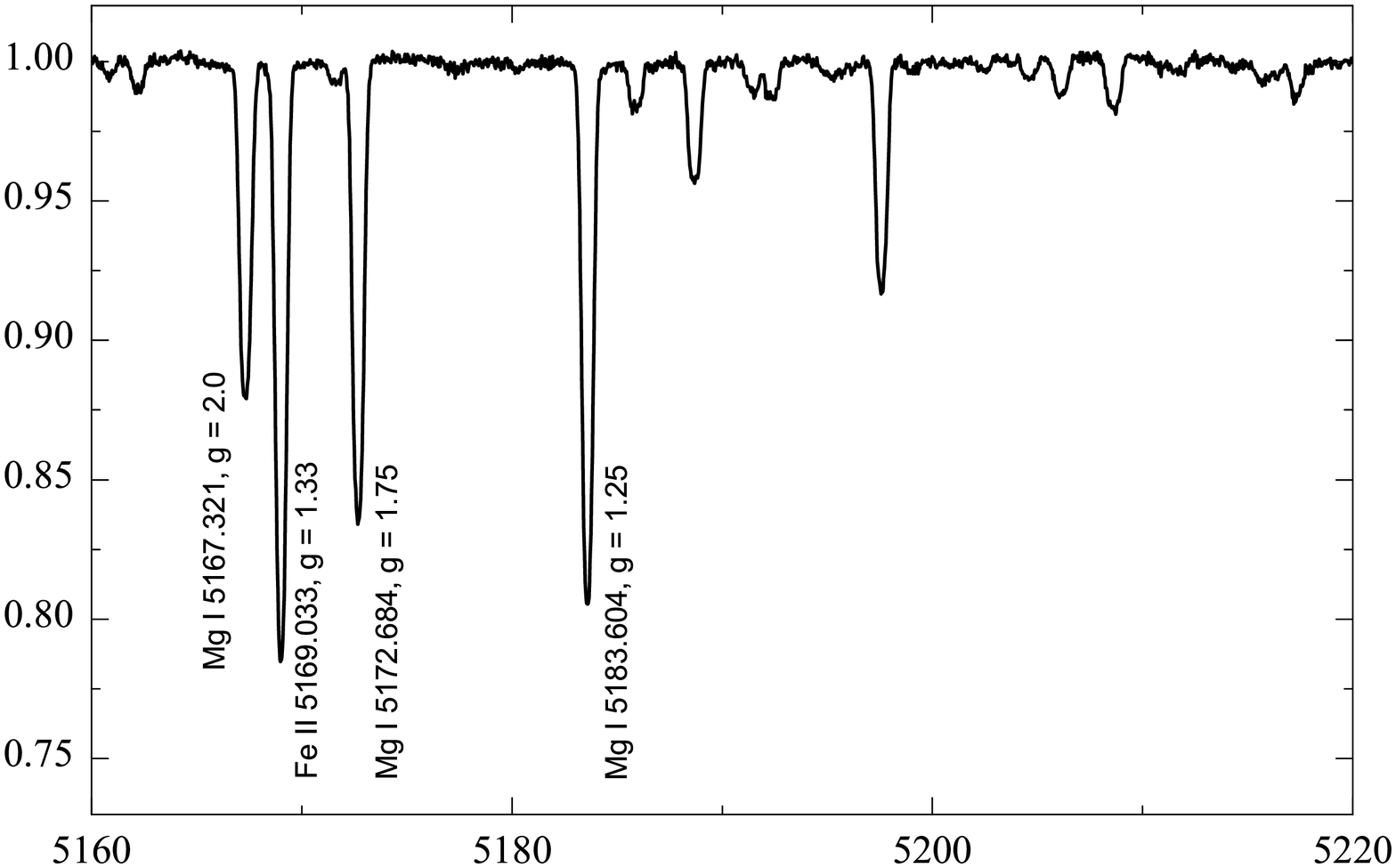}
 
}
\vspace{-5mm}
\caption[]{The normalized observed spectrum of Vega, 
$g$ -- effective Lande factor of the spectral lines.
}
\label{spectra}
\end{center}
\end{figure}

High-accuracy spectropolarimetric study of Vega has been performed in the spectral region 5170 \AA ~during 37 nights from 1997 to 2012. The star was observed using coude spectrograph at the 2.6-m Shajn telescope mounted at the Crimean Astrophysical Observatory (CrAO, Ukraine). Over 2700 circularly polarized high-resolved spectra with total signal-to-noise ratio per pixel about 37500 were recorded. The resolving power of the spectra is approximately 25000. The effective magnetic field was calculated using the technique which is described in detail by Plachinda \cite{plach04}, and Butkovskaya \& Plachinda \cite{butk07}. We calculated Zeeman splitting for each single line marked at Figure~\ref{spectra}. For each single line the 1353 single values of the longitudinal magnetic field were obtained. Then we calculated 1353 mean longitudinal magnetic field values averaged by all four lines. To ensure an optimal data quality  we used Law of Large Numbers (LLN). Due to LLN magnetic field values which exceeded the mean value by 3$\sigma$ were eliminated from our dataset. The final time-string consists of the 1312 mean longitudinal magnetic field measurements.

\section{Results}\label{results}

\begin{figure}[!t]
\begin{center}
\hbox{
 \includegraphics[width=10cm]{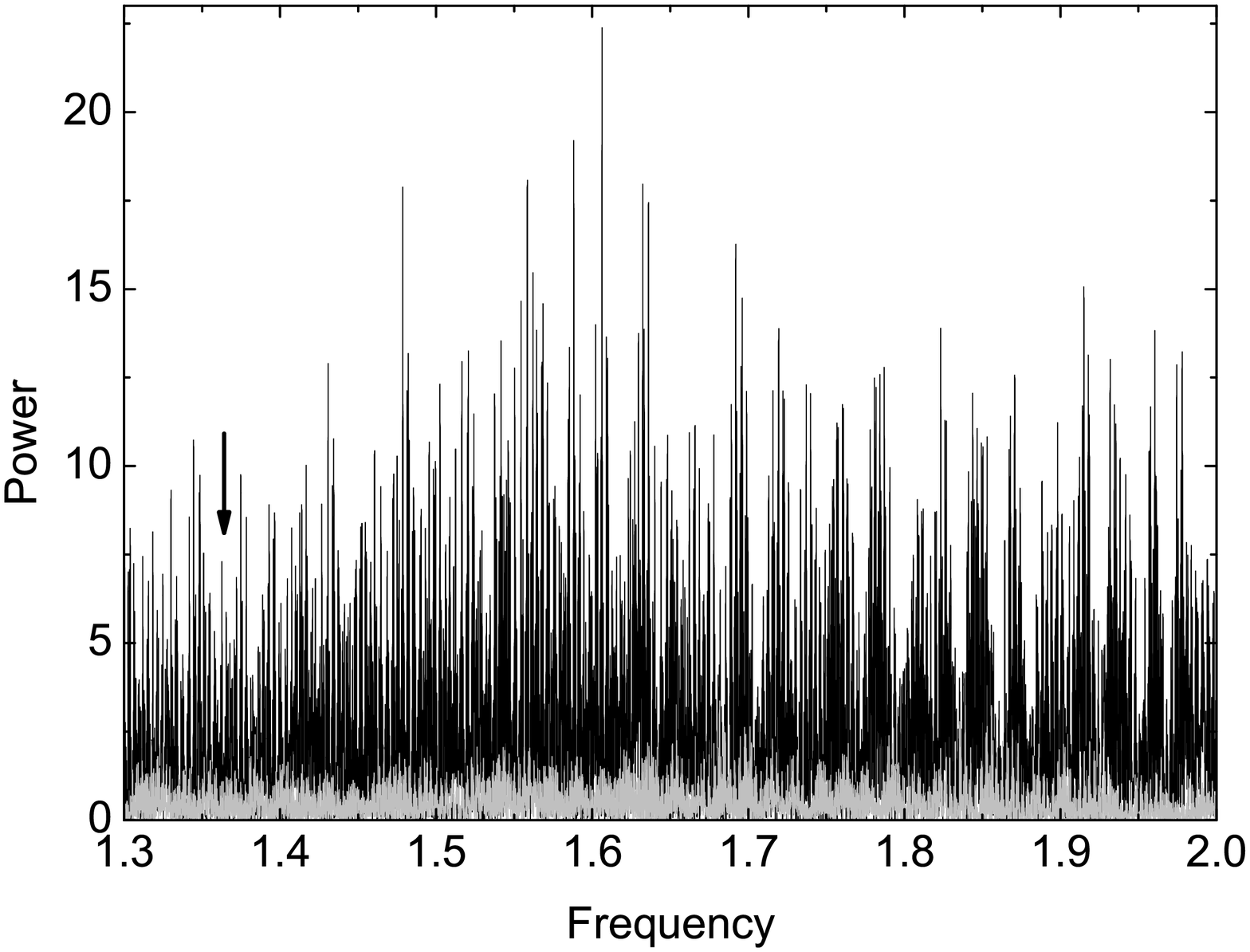}
 
}
\vspace{-5mm}
\caption[]{The power spectra of the longitudinal magnetic field (\textit{black}), and the 'null' field which characterizes the spurious magnetic signal (\textit{gray}). The maximal peak corresponds to the frequency 1.6062959 $d^{-1}$. The arrow marks the frequency 1.3663 $d^{-1}$ which corresponds to rotation period proposed by Petit et al. \cite{petit10}.

}
\label{power}
\end{center}
\end{figure}

Search for a periodicity in the longitudinal magnetic field data was performed using the Period04 code. The range of the analyzed frequencies encompasses the rotation periods already proposed in the literature (Aufdenberg et al. \cite{aufdenberg06}, Takeda et al. \cite{takeda08}, Hill et al. \cite{hill10}, Petit et al. \cite{petit10}). The power spectrum of the longitudinal magnetic field (Figure~\ref{power}, black line) revealed the most prominent peak at frequency 1.6062959 $d^{-1}$ (signal-to-noise is 3.07) which corresponds to the period 0.62255 d. The period is very close to the rotation period 0.622 d estimated by Hill et al. \cite{hill10}. The power spectrum of the 'null' field (Figure~\ref{power}, gray line) revealed no any significant peaks in this frequency range.

\begin{figure}[!t]
\begin{center}
\hbox{
 \includegraphics[width=10cm]{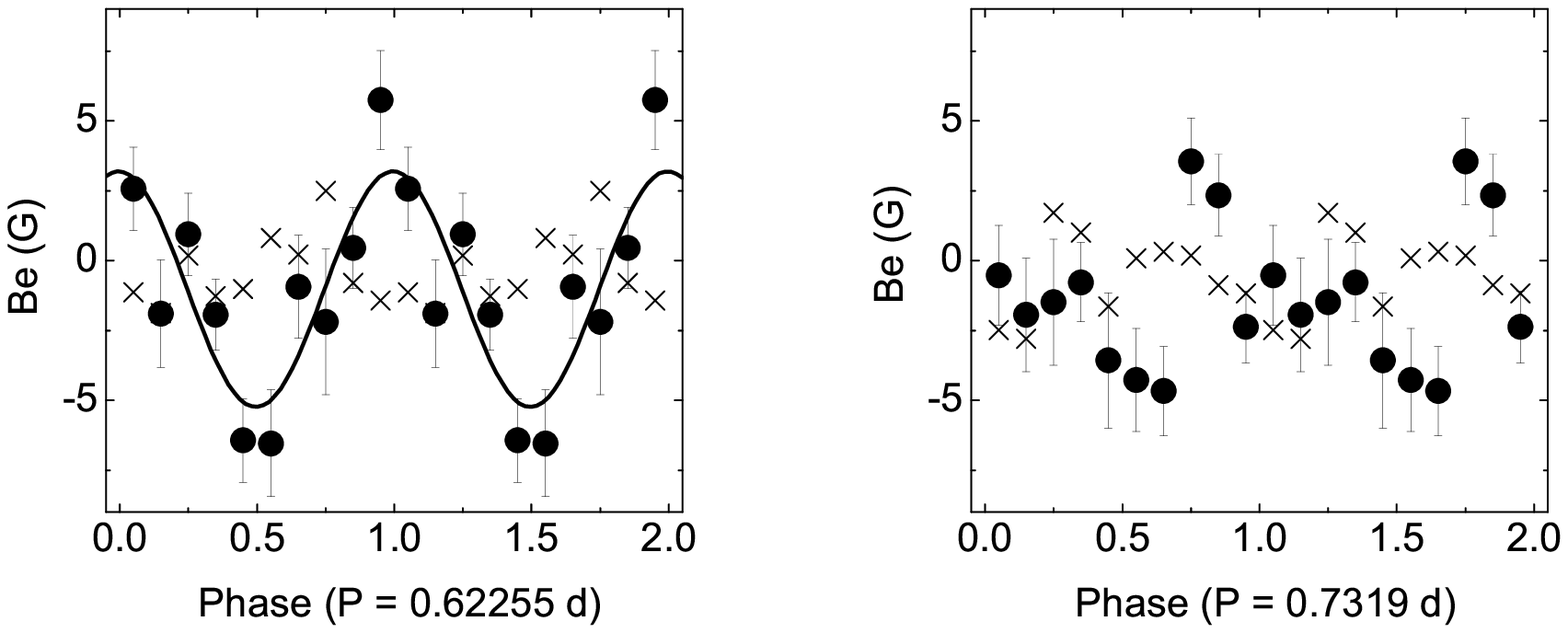}
 
}
\vspace{-25mm}
\caption[]{\textbf{Left}: the longitudinal magnetic field (\textit{black circles}) and 'null' field (\textit{crosses}) of Vega phased with the rotation period 0.62255 d, and averaged within 10 bins. \textbf{Right}: the longitudinal magnetic field (\textit{black circles}) and 'null' field (\textit{crosses}) phased with the rotation period 0.7319 d (Petit et al. \cite{petit10}), and averaged within 10 bins. Least-square sinusoidal fit is shown by strong line.

}
\label{field}
\end{center}
\end{figure}

We adopt the Julian date $JD$ = 2450658.427 as phase origin, and phased the longitudinal magnetic field of Vega with the 0.62255-day period. In the left panel of Figure~\ref{field} the longitudinal magnetic field of Vega averaged within 10 bins is presented. The estimated by Fisher test statistical reliability of the magnetic field curve vs phased 'null' field is 99.8\%.

In attempt to reconstruct a relevant topology of the surface field on Vega, Petit et al. \cite{petit10} calculated the set of magnetic maps, assuming for each map a different value for the rotation period. They adopt a value of 0.7319 d for the rotation period, and choose the Julian date $JD$ = 2454101.5 as phase origin. We tested the agreement of our long-term magnetic field measurements with the period proposed by Petit et al. \cite{petit10}. The frequency 1.3663 $d^{-1}$ (signal-to-noise is 1.28) is marked in Figure~\ref{power} by  arrow. We phased the longitudinal magnetic field of Vega with the 0.7319-day period. In the right panel of Figure~\ref{field} the longitudinal magnetic field of Vega averaged within 10 bins is presented. The estimated by Fisher test statistical reliability of the magnetic field curve vs phased 'null' field is 91.5\%.

Our long-term longitudinal magnetic field measurements do not confirm the rotation period proposed by Petit et al. \cite{petit10}. On the other hand, the different periods can be caused by different bulk of spectral lines used by us and by Petit et al. \cite{petit10}. The lines can be formed in the different physical conditions at different latitudes.

\bigskip
{\it Acknowledgements.} The author thanks S. Plachinda for useful discussions, and D. Baklanova and S. Plachinda for the help in observations.


\begin{thebibliography}{99}

\bibitem{alina11}
{Alina, D. et al.} 2011, AN, 332, 943

\bibitem{aufdenberg06}
{Aufdenberg, J. P. et al.} 2006, ApJ, 645, 664

\bibitem{butk07}
{Butkovskaya V., Plachinda S.} 2007, A\&A, 469, 1069

\bibitem{butk11}
{Butkovskaya, V., Plachinda, S., Valyavin, G., Baklanova, D., \& Lee, B.-C.} 2011, AN, 332, 956

\bibitem{gulliver94}
{Gulliver, A. F., Hill, G., \& Adelman, S. J.} 1994, ApJ, 429, L81

\bibitem{hill10}
{Hill, G., Gulliver, A. F., \& Adelman, S. J.} 2010, ApJ, 712, 250

\bibitem{lignieres09}
{Lignieres, F., Petit, P., Bohm, T. \& Auriere, M.} 2009, A\&A, 500, L41

\bibitem{petit10}
{Petit, P. et al.} 2010, A\&A, 523, id.A41

\bibitem{plach04}
{Plachinda, S. I.} 2004, in Photopolarimetry in Remote Sensing, ed. G. Videen, Ya. S. Yatskiv, 
\& M. I. Mishchenko (Kluwer Acad. Publ.), 351

\bibitem{takeda08}
{Takeda,Y., Kawanomoto, S., and Ohishi N.} 2008, ApJ, 678, 446

\bibitem{vasil89}
{Vasil'ev, I.A. et al.} 1989, IBVS 3308

\end{thebibliography}
\end{document}